\documentclass[aps,prl,reprint,shortbibliography,superscriptaddress]{revtex4-1}
\usepackage{ragged2e}
\usepackage{setspace}
\usepackage{amsmath}
\usepackage{graphicx}
\usepackage[nearskip,margin = 0pt]{subfig}
\usepackage{subfig}

\usepackage{verbatim}
\usepackage{amsfonts}
\usepackage{amssymb}
\usepackage{epstopdf} 
\usepackage{xcolor}
\DeclareGraphicsExtensions{.pdf,.eps,.png,.jpg,.mps} 
\usepackage{hyperref}
\hypersetup{
    colorlinks=true,
    linkcolor=blue,
    citecolor=blue,
    filecolor=blue,      
    urlcolor=blue,
}

\begin{document}

\title{Self-injection locking dynamics with Raman actions in AlN microresonators}

\author{Yulei Ding}
\thanks{These authors contribute equally to the work}
\affiliation{State Key Laboratory of Precision Measurement Technology and Instruments, Department of Precision Instruments, Tsinghua University, Beijing 100084, China}

\author{Yifei Wang}%
\thanks{These authors contribute equally to the work}
\affiliation{State Key Laboratory of Precision Measurement Technology and Instruments, Department of Precision Instruments, Tsinghua University, Beijing 100084, China}

\author{Shunyu Yao}%
\affiliation{State Key Laboratory of Precision Measurement Technology and Instruments, Department of Precision Instruments, Tsinghua University, Beijing 100084, China}

\author{Yanan Guo}%
\affiliation{Center for Materials Science and Optoelectronics Engineering, University of the Chinese Academy of Sciences, Beijing 100049, China}
\affiliation{Research and Development Center for Solid State Lighting, Institute of Semiconductors, Chinese Academy of Sciences, Beijing 100083, China}

\author{Jianchang Yan}%
\affiliation{Center for Materials Science and Optoelectronics Engineering, University of the Chinese Academy of Sciences, Beijing 100049, China}
\affiliation{Research and Development Center for Solid State Lighting, Institute of Semiconductors, Chinese Academy of Sciences, Beijing 100083, China}

\author{Junxi Wang}%
\affiliation{Center for Materials Science and Optoelectronics Engineering, University of the Chinese Academy of Sciences, Beijing 100049, China}
\affiliation{Research and Development Center for Solid State Lighting, Institute of Semiconductors, Chinese Academy of Sciences, Beijing 100083, China}

\author{Changxi Yang}%
\affiliation{State Key Laboratory of Precision Measurement Technology and Instruments, Department of Precision Instruments, Tsinghua University, Beijing 100084, China}

\author{Chengying Bao}
\email{cbao@tsinghua.edu.cn}
\affiliation{State Key Laboratory of Precision Measurement Technology and Instruments, Department of Precision Instruments, Tsinghua University, Beijing 100084, China}

\begin{abstract}

Self-injection locking (SIL) of semiconductor lasers to on-chip microcavities enables significant laser noise purification and diverse nonlinear optical actions. Realizing nonlinear SIL in new material platforms is essential for advancing photonic integrated circuits. Here, we demonstrate nonlinear SIL in AlN microcavities that generates stimulated Raman lasers (SRLs) and microcombs. We achieve SRL emission with an output power exceeding 10 mW and a fundamental linewidth below 70 Hz in the 1750 nm band. 
The Kerr effect further mediates stimulated emissions at the 2nd-Stokes and anti-Stokes frequencies. Multi-time-scale thermal relaxations during turnkey SIL enable GHz-level frequency sweeps of the SRL and pump. 
Raman actions also render a Stokes platicon microcomb state with co-emission in the pump and Stokes bands. Hybrid-integrated crystalline microresonators can be a versatile platform to investigate nonlinear photon-phonon interactions.
\end{abstract}

\maketitle

\textit{Introduction.-} 
Rayleigh backscattering in microresonators couples the forward pumped mode with the backward orbiting mode \cite{Vahala_OL2002modal}. Collecting backscattered light back into the pump laser can result in frequency locking to the microresonator resonance, which is  known as self-injection locking (SIL). SIL leads to 
dramatic laser noise reduction in a compact system architecture \cite{Matsko_NC2015ultralow,Gorodetsky_OE2017self,kondratiev2023recent}. 
When pump power is sufficiently high, nonlinear actions in the microresonators can be activated, enabling highly coherent, turnkey microcombs \cite{Matsko_NC2015,Gorodetsky_NP2018narrow,Bowers_Nature2020integrated,Bilenko_NC2021dynamics,Herr_NP2024synthetic}. Initially, SIL to microresonators predominately focused on crystalline whispering gallery mode microresonators \cite{Matsko_NC2015,Matsko_NC2015ultralow,Gorodetsky_NP2018narrow}. Recently, SIL to chip-integrated high-Q microresonators has witnessed great success via either hybrid integration (separate chips integrated by butt-coupling) \cite{Lipson_Nature2018battery,Kippenberg_NC2019electrically,Bowers_Nature2020integrated,Bowers_NP2021hertz,lihachev2022platicon,Vahala_PRA2022self,Herr_NP2024synthetic,Vahala_Science2024multimodality,Vahala_OL2021reaching,Bowers_SA1Hz} or monolithic integration \cite{Bowers_Science2021laser,Bowers_Nature2023}. 
These progresses have brought us closer to synthesizing ultralow noise lasers \cite{Bowers_NP2021hertz,Vahala_OL2021reaching,Bowers_SA1Hz,Bowers_Science2021laser,Bowers_Nature2023}, microcombs \cite{Lipson_Nature2018battery,Kippenberg_NC2019electrically,Bowers_Science2021laser,Bowers_NP2021hertz,lihachev2022platicon,Vahala_PRA2022self,Herr_NP2024synthetic,Vahala_Science2024multimodality,Bowers_Nature2020integrated} and microwaves \cite{Diddams_Nature2024photonic,Bowers_Nature2023} in photonic integrated circuits (PICs).

Silicon nitride (Si$_3$N$_4$) microresonators (either loosely or tightly confined) are the main workhorse in this endeavor \cite{Lipson_Nature2018battery,Kippenberg_NC2019electrically,Bowers_Nature2020integrated,Bowers_NP2021hertz,lihachev2022platicon,Vahala_PRA2022self,Bowers_Science2021laser,Bowers_Nature2023,Vahala_OL2021reaching,Herr_NP2024synthetic}. Other integrated high-Q platforms, including silica \cite{Vahala_NP2018bridging}, lithium niobate (LN) \cite{LinQ_Optica2019self}, tantala \cite{Papp_Optica2021tantala}, AlGaAs \cite{Bowers_NC2020ultra}, and AlN \cite{Tang_AOP2023aluminum}, have also show great prospects in nonlinear photonic applications. Unlocking the potential of different material platforms is critical to the development of PICs. 
In addition to Kerr microcombs, nonlinear SIL has also enabled second harmonic generation (SHG) in LN \cite{LinQ_LPR2023self} and optically poled Si$_3$N$_4$ \cite{Vahala_Optica2023high,Bres_LSA2023chip} microcavities. More recently, SIL-based stimulated Brillouin laser (SBL) has also been demonstrated in coupled Si$_3$N$_4$ microcavities \cite{Vahala_Science2024multimodality}. As a counterpart of SBL, stimulated Raman laser (SRL) in microresonators via SIL has not been demonstrated, to our knowledge. SRLs shift the lasing frequency by the optical phonon frequency, reaching spectral regions far beyond pumps. 
Integrated crystalline material platforms (e.g., AlN \cite{SunCZ_Optica2016} and LN \cite{Loncar_LSA2020raman}) featuring narrow and strong Raman gain spectra are well-suited for SIL-SRL generation. Together with the Kerr effect, Raman actions can initiate comb generation in the Raman gain band, thus extending the operation bandwidth of microcombs \cite{Vahala_NP2017Stokes,Chembo_PRA2015spatiotemporal}. 

Here, we report a nonlinear SIL system with Raman actions using a hybrid-integrated AlN-on-sapphire platform \cite{Tang_AOP2023aluminum,Tang_OL2013optical,Bao_OL2022_Raman,Bao_OL2021self,liu2023mitigating}. A high power distributed feedback (DFB) laser SIL to high-Q AlN microresonators was used to generate SRLs and dark pulse (platicon) microcombs \cite{Bowers_NP2021hertz,lihachev2022platicon,Weiner_NP2015mode}. The SRL has a fundamental linewidth less than 70 Hz. SIL also enables emissions at the 2nd-Stokes (2S) and anti-Stokes (AS) frequencies. 
Thermal relaxations were seldom investigated in SIL, as SIL-microcombs are free from thermal instability \cite{Bowers_Nature2020integrated,Bowers_NP2021hertz}. We observe transient thermal relaxations occurring in time-scale spanning from sub-$\mu$s to tens-ms enable frequency sweep over several GHz for both the pump and the SRL in turnkey SIL. These hybrid-integrated frequency swept lasers may be used for two-color LiDARs \cite{Leaird_OL2024wavelength}. 
In addition, platicon microcombs have not been demonstrated in AlN microresonators, as their generation encounters challenges including narrow existence range and fast thermal instability (an order of magnitude faster than Si$_3$N$_4$ microcavities)
\cite{Yang_PRA2020generation,liu2023mitigating}. SIL enabled the generation of such an AlN platicon microcomb. We also observed a `Stokes platicon' state with co-emission of microcombs in the pump and the Raman bands, in analogy to the silica bright Stoke solitons \cite{Vahala_NP2017Stokes}. 

\begin{figure}[t!]
\captionsetup{singlelinecheck=no, justification = RaggedRight}
\includegraphics[width=\linewidth]{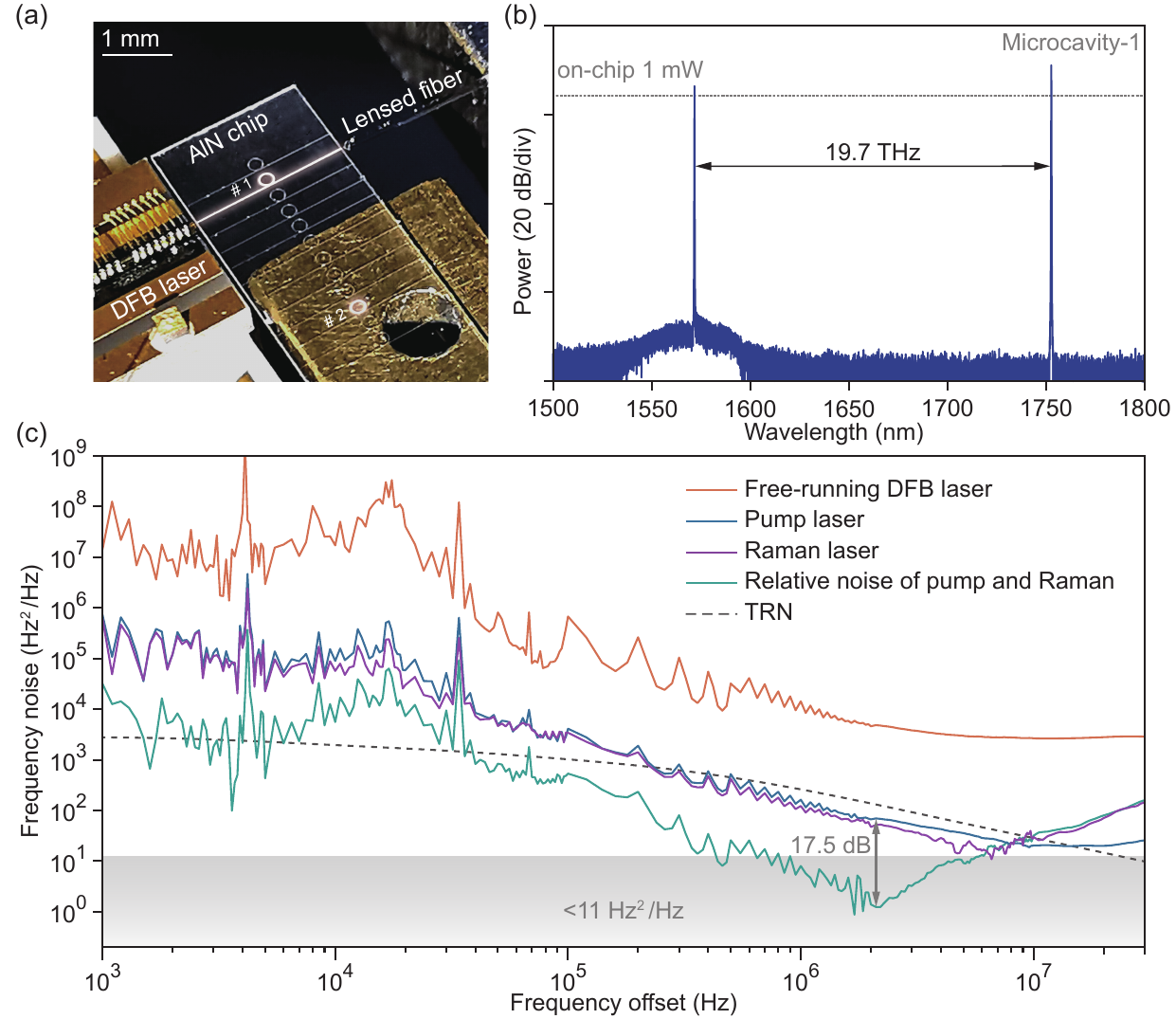}
\caption{
(a) Picture of a DFB laser butt-coupled to the AlN photonic chip. The false color highlights the used microcavities. (b) Optical spectrum of the SIL-SRL. 
(c) Single sideband frequency noise spectra of the free-running pump (yellow), SIL pump (blue), and SRL (purple). The gray dashed line is the simulated thermo-refractive noise (TRN) for the AlN microcavity, and the green curve is the relative frequency noise between the pump and the SRL.} 
\label{fig1}
\end{figure}

\textit{Raman lasing via SIL.-} The AlN microresonators have the same design and fabrication process with our previous work \cite{Bao_OL2022_Raman,Bao_OL2021self,liu2023mitigating}. They have a diameter of 200 $\mu$m and a cross-section of 3.5$\times$1.1 $\mu$m, resulting in normal dispersion for the TE$_{\rm 00}$ mode and anomalous dispersion for the TE$_{\rm 10}$ mode. 
We first pumped a TE$_{\rm 00}$ mode of microcavity-1 via butt-coupling (Fig. \ref{fig1}(a)). The laser-to-chip coupling loss is about 5 dB and the pumped mode has an intrinsic (loaded) Q-factor of 2.0 (1.1)$\times10^6$. The mode has a relatively strong backscattering, as evidenced by mode splitting (Supplementary Sec. 1 \cite{Supplement}). By tuning the laser into the resonance, SIL can be realized. When the pump power exceeds the threshold (on-chip power $\sim$13 mW), SRL starts to emit. The large optical phonon frequency of AlN creates SRL at 1753 nm when pumping at 1572 nm (Fig. \ref{fig1}(b)). The on-chip SRL power reaches 4.8 mW with an on-chip pump power of 30 mW. Higher power is feasible using over-coupled devices \cite{Bao_OL2022_Raman}.

\begin{figure}[t!]
\captionsetup{singlelinecheck=no, justification = RaggedRight}
\includegraphics[width=\linewidth]{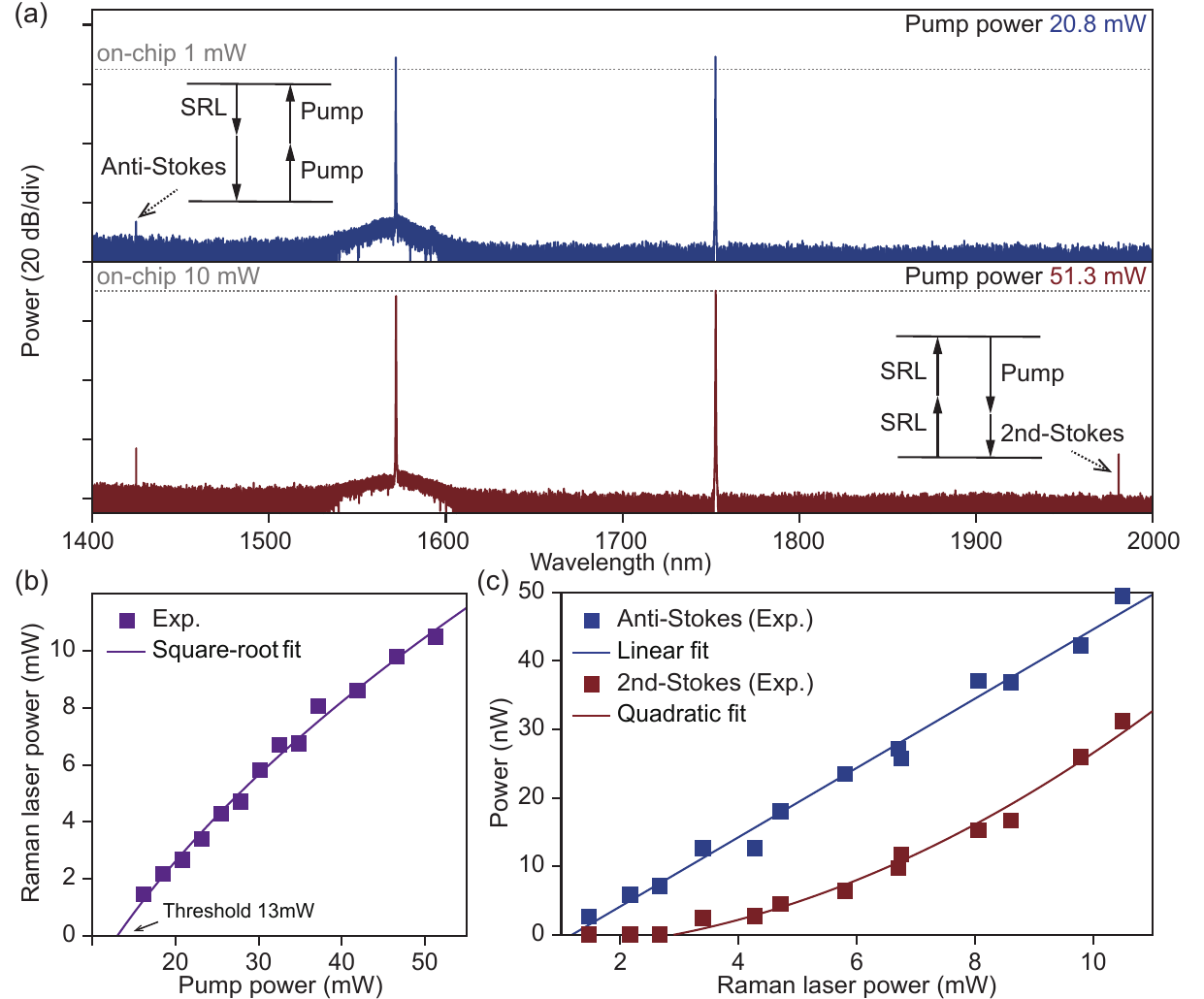}
\caption{
(a) Optical spectra when pumping the AlN microcavity with a pump power of 20.8 mW and 51.3 mW, respectively. The diagrams illustrate the generation dynamics for the anti-Stokes (AS) and 2nd-Stokes (2S) emissions via the Kerr effect. (b) Measured SRL power and its square-root fit. (c) Measured AS power changes for the AS and 2S emissions and their fits.}
\label{fig1Apower}
\end{figure}

\begin{figure}[t]
\captionsetup{singlelinecheck=no, justification = RaggedRight}
\includegraphics[width=\linewidth]{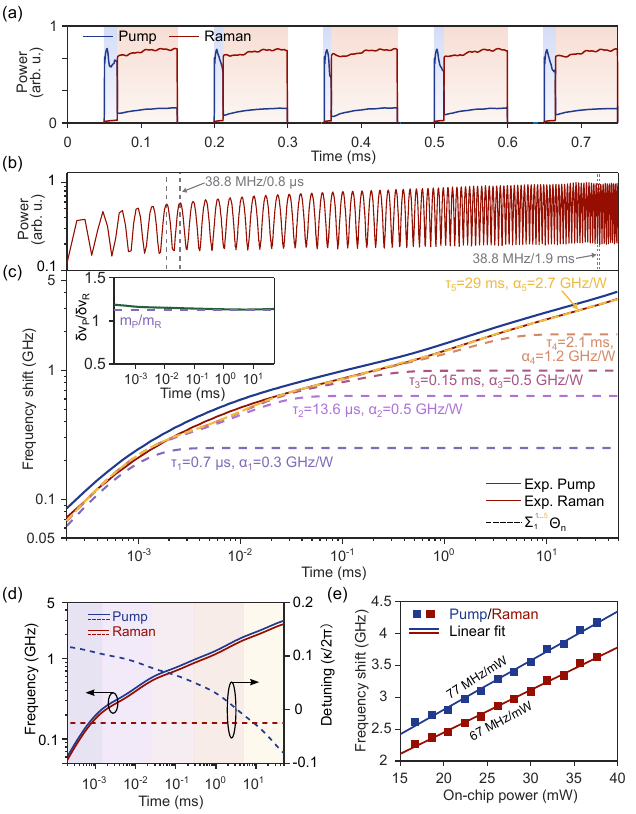}
\caption{(a) Measured pump and SRL power when turning on/off the pump laser by square waves. 
(b) SRL interference signal measured by a Mach-Zehnder interfereometer. (c) Frequency change of the pump and SRL during the turnkey operation. Five thermal relaxation processes are used to fit the measured frequency change. The inset shows the ratio between $\delta\nu_{\rm R}$ and $\delta\nu_{\rm P}$, which is close to $m_{\rm P}/m_{\rm R}$. (d) Simulated frequency change and effective detunings for the pump and SRL. (e) Measured magnitude of frequency change of for the pump and the SRL under different pump powers. The solid lines corresponds to linear fits.}
\label{fig3}
\end{figure}

\begin{figure}[t]
\centering
\captionsetup{singlelinecheck=no, justification = RaggedRight}
\includegraphics[width=\linewidth]{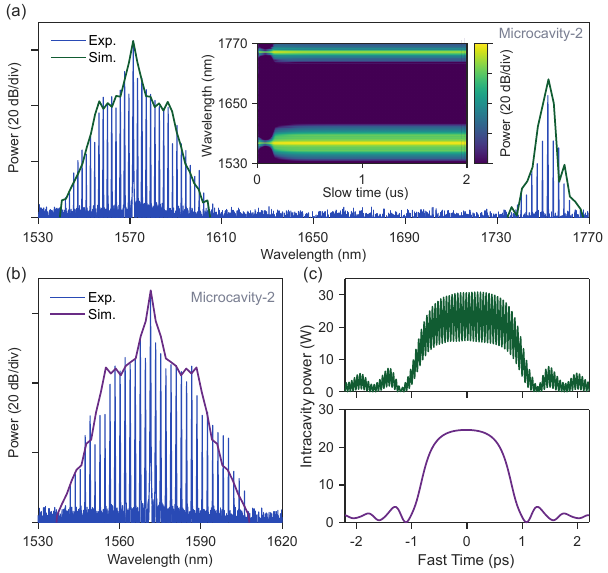}
\caption{(a) Measured and simulated platicon microcomb with comb lines emerging in the Raman band. The inset shows the simulated comb generation dynamics. (b) A platicon microcomb without the Raman band lines generated by tuning the laser-to-chip phase delay $\phi_B$. (c) Simulated platicon waveforms with (upper) and without (bottom) the Raman band lines. }
\label{fig4}
\end{figure}


We then measured the frequency noise of the pump and the SRL with an power of 4.8 mW using the delayed self-heterodyne method \cite{Bowers_NP2021hertz,Bao_OL2022_Raman,Vahala_OE2022correlated} (Fig. \ref{fig1}(c)). 
SIL suppressed the pump frequency noise by about 19 dB (23 dB) at an offset frequency of 3 MHz (10 kHz) from the free-running DFB laser. Further noise reduction is hindered by the small mode volume and the resulted high thermo-refractive noise (TRN) limit (dashed curve in Fig. \ref{fig1}(c) is the simulated TRN limit) \cite{Bowers_NP2021hertz,Vahala_OL2021reaching}. The frequency noise of the SRL mimics the SIL pump wave. Since the used photodetector has weak response in the 1700 nm band, the noise floor causes the measured frequency noise to rise above 7 MHz. The fundamental frequency noise of the SRL should be less than 11 Hz$^2$/Hz (i.e., a fundamental linewidth of 70 Hz), which is comparable to the conventional pumping case \cite{Bao_OL2022_Raman}. Narrower linewidth can be achieved by increasing the Q-factor or the SRL power. 

With the retrieved phase of the pump and the SRL waves, we analyzed the relative frequency noise between them (green curve in Fig. \ref{fig1}(c)). Due to common noise rejection, the relative frequency noise is lower than the pump and SRL frequency noise. However, the common noise rejection is not perfect, as mode numbers for the pump and the SRL differ considerably. Based on the simulated effective refractive index, mode numbers for the pump frequency ($\nu_{\rm P}$) and the SRL frequency ($\nu_{\rm R}$) were estimated to be $m_{\rm P(R)}$=806(718). 
The theoretical common noise rejection ratio limit is $m_{\rm R}^2/(m_{\rm P}-m_{\rm R})^2$ \cite{Loh_Optica2024ultralow}, which is $\sim$18 dB for our sample. The measured frequency noise reduction between the relative and absolute noise is close to this limit around 2 MHz. This is consistent with the simulation that the frequency noise at these offset frequencies are limited by the TRN. It also suggests that the pump and SRL are accommodated by the same spatial mode family. The relative noise reduction is slightly lower than this limit for lower offset frequencies, which we believe may result from the different intensity noise response for the pump and SRL waves via the Kerr effect (see discussions in following Fig. \ref{fig3} and Eqs. \ref{eq1SRL}, \ref{eq2Pump}).



\textit{2nd-Stokes and anti-Stokes emissions.-} We then measured the SIL-SRL power change by varying the pump power. Notably, stimulated emissions at AS and 2S frequencies were also observed (Fig. \ref{fig1Apower}(a)). The AS emission at 1425 nm appeared prior to the 2S emission at 1981 nm (see 
Fig. \ref{fig1Apower}(a)). These emissions are mediated by the Kerr effect rather than the Raman actions, as evidenced by the measured power change. The SIL-SRL power has a square-root scaling with the pump power (Fig. \ref{fig1Apower}(b)), which is the same with conventionally pumped SRLs \cite{kippenberg2004Raman,Bao_OL2022_Raman}. The highest SRL power can reach 10.5 mW with a pump power of 51.3 mW. The AS and 2S emissions have a linear and a quadratic scaling with the SRL power, respectively (Fig. \ref{fig1Apower}(c)). For the AS lasing, the linear power change corresponds to a four-wave-mixing (FWM) process using 2 photons at $\nu_{\rm P}$ as the pump and a photon at $\nu_{\rm R}$ as the idler. The quadratic-scaling 2S lasing is rooted in FWM using 2 photons at $\nu_{\rm R}$ as the pump and a photon at $\nu_{\rm P}$ as the signal (inset of Fig. \ref{fig1Apower}(a)). 
Thus, SIL in AlN microresonators can also extend the lasing spectrum into the 2000 nm band for chip-integrated devices, which is important to enhance the communication capacity beyond the telecom C-band \cite{Eggleton_JOP2016roadmap,gunning2018key}. Numerical simulations show similar power changes for the SRL, AS and 2S emissions (Supplementary Sec. 2 \cite{Supplement}). The 2nd cascaded SRL threshold has not been reached for the measured SRL power (once it is reached, the SRL power will be clamped \cite{kippenberg2004Raman,Supplement}).

\textit{Frequency sweep in turnkey SIL.-} As a feature of SIL systems, SRL can be initiated immediately after turning on the pump laser (i.e., turnkey operation \cite{Bowers_Nature2020integrated}). Fig. \ref{fig3}(a) shows the measured power changes for the pump and SRL, when switching on/off the pump by periodic square waves. Once the pump frequency approached the resonance (transition from blue to red shaded boxes), it was SIL to the microcavity and the output pump power decreased abruptly. Meanwhile, the SRL power rose abruptly to steady states. We then measured the frequency changes for both the pump and the SRL during turnkey SIL by Mach-Zehdner interferometers (MZIs). An example of the MZI output for the SRL is plotted in Fig. \ref{fig3}(b), showing decreasing chirp rates. The real-time change of the laser frequency was then retrieved in Fig. \ref{fig3}(c). The SRL and pump had frequency changes of 3.6 GHz and 4.1 GHz, respectively. The amplitudes are slightly larger than lasers SIL to piezo-electrically or elctro-optically tuned microcavities, that have been used for frequency modulated continuous wave (FMCW) LiDARs \cite{Kippenberg_NC2022low,Kippenberg_Nature2023ultrafast}. After correcting the frequency sweep nonlinearity \cite{Bao_arXiv2024microcomb}, the system can be used for dual-color FMCW LiDARs with speed penalties.

Due to the mode-pulling effect, SRL has a detuning from the resonance as  $\nu_{\rm R0}-\nu_{\rm R}=-(\nu_{\rm P}-f_{\rm R}-\nu_{\rm R})\kappa/\Gamma_{\rm R}$, where $\nu_{\rm R0}$, $f_{\rm R}$, $\kappa/2\pi$, $\Gamma_{\rm R}/2\pi$ are the cavity resonance frequency for the Raman lasing mode, phonon frequency, cavity linewidth and linewidth of the Raman gain spectrum, respectively \cite{Bennett1962gaseous,Vahala_OE2012SBL,Supplement}. 
As $\Gamma_{\rm R}$ is about three orders of magnitude larger than $\kappa$ for AlN microcavities \cite{Bao_OL2022_Raman}, the measured SRL frequency change follows the resonance shift as (with AS and 2S powers neglected),

\begin{equation}
2\pi\delta \nu_{\rm R}(T)=\Theta(T)+\gamma P_{\rm R} v_g+2\gamma (P_{\rm P} +P_{\rm B}) v_g,
\label{eq1SRL}
\end{equation}
where $T$ is the measurement time, $\Theta$ is the resonance shift due to thermal effects for the SRL mode, $\gamma$ is the Kerr nonlinearity coefficient, $v_g$ is the group velocity of the microcavity, $P_{\rm R(P)}$ is the intracavity SRL (pump) power and $P_{\rm B}$ is the backscatterred power including both the SRL and pump. These powers were estimated to be $P_{\rm P}$=1.5 W, $P_{\rm R}$=1.4 W, and $P_{\rm B}$=2.0 W (Supplementary Sec. 3 \cite{Supplement}). In our definition, a positive $\delta \nu_{\rm R}$ stands for red-shift of the frequency. Similarly, the pump frequency change can be written as,

\begin{equation}
2\pi\delta \nu_{\rm P}(T)=\frac{m_{\rm P}}{m_{\rm R}}\Theta(T)+2\gamma (P_{\rm R}+P_{\rm B}) v_g+\gamma P_{\rm P} v_g + \delta\omega_{\rm P}(T),
\label{eq2Pump}
\end{equation}
where $\delta\omega_{\rm P}$ is the frequency detuning change in SIL for the pump. The Kerr effect shifts the resonance differently for the pump and the SRL mode. We compare $\delta \nu_{\rm P}/\delta \nu_{\rm R}$ in the inset of Fig. \ref{fig3}(c), which is close to $m_{\rm P}/m_{\rm R}$. It suggests the frequency change is dominated by thermal effects. In turnkey operation, $P_{\rm R}$ and $P_{\rm P}$ become steady quickly, and $\Theta$ can be written as,

\begin{equation}
\Theta(T) =\sum\nolimits_n \Theta_n(T)=\sum\nolimits_n \alpha_n P_{t}(1-e^{-T/\tau_n}),
\label{eq1Thermal}
\end{equation}
where $\Theta_n$ is the $n$th thermal relaxation process with $\alpha_n$ and $\tau_n$ being its strength and relaxation time, and $P_{t}=P_{\rm R}+P_{\rm P}+P_{\rm B}$. The actual thermal relaxation can be quite complicated, but we found the measured $\delta \nu_{\rm R}(T)$ can be reasonably fitted by five relaxation processes with $\tau_n$ ranging from 0.7 $\mu$s to 29 ms (Fig. \ref{fig3}(c)). The dashed curves show the contributions from each $\Theta_n$ with $\tau_n$ and $\alpha_n$ listed in Fig. \ref{fig3}(c). The retrieved $\sum \alpha_n$ is 60 times of $\gamma v_g$ (simulated to be about 86 MHz/W). $\alpha_n$ derived in independent measurements of thermal broadening of the resonance under different laser scanning speeds are consistent with the fitted $\alpha_n$ (Supplementary Sec. 4 \cite{Supplement}).

Simulation shows similar frequency change trend and magnitude (Fig. \ref{fig3}(d), see also Supplementary Sec. 5 \cite{Supplement}). Simulations allow us to compute the effective detuning for the SRL as $\delta\nu_{\rm R}-[\Theta+\gamma v_g (P_{\rm R}+2P_{\rm P}+2P_{\rm B})]/2\pi$ (positive for red-detuning). It shows that the effective detuning for the SRL stays nearly constant in turnkey SIL (red dashed line Fig. \ref{fig3}(d)) and $\nu_{\rm R}$ follows $\nu_{\rm R0}$. Similarly, the effective detuning for the pump was analyzed as $\delta\nu_{\rm P}-[m_{\rm P}\Theta/m_{\rm R}+\gamma v_g (P_{\rm P}+2P_{\rm R}+2P_{\rm B})]/2\pi$ (without accounting for the resonance shift due to backscattering), which changes about 0.2$\kappa/2\pi$ in SIL. 

To further validate that $\delta \nu_{\rm R}$ and $\delta \nu_{\rm P}$ are dominated by $\Theta$, we measured their magnitudes under different pump powers (Fig. \ref{fig3}(e)). Both magnitudes increase linearly with increasing pump powers. The fitted slopes are 77 MHz/mW for the pump and 67 MHz/mW for the SRL, respectively. Their ratio is 1.15, close to $m_{\rm P}/m_{\rm R}$=1.12. Once SRL is generated, the intracavity pump power $P_{\rm P}$ in Eqs. \ref{eq1SRL}, \ref{eq2Pump} is fixed at the intracavity threshold and independent from external pump power \cite{Bao_OL2022_Raman,kippenberg2004Raman}. Since $P_{\rm R}$ impacts $\delta\nu_{\rm P}$ stronger than $\delta\nu_{\rm R}$, the measured ratio is slightly higher than $m_{\rm P}/m_{\rm R}$. 

\textit{Raman actions in microcomb generation.-} Impact of Raman actions on microcombs was also observed in the hybrid-integrated AlN platform. Here, a TE$_{00}$ mode of  microcavity-2 in Fig. \ref{fig1}(a), whose intrinsic (loaded) Q-factor is 2.9 (1.4)$\times10^6$, was pumped at 1571 nm with an on-chip power of $\sim$55 mW. By adjusting the phase delay ($\phi_{\rm B}$) between the DFB laser and the AlN microcavity, a platicon microcomb with over 30 lines spaced by $\sim$230 GHz was generated. Different from SIL platicon microcombs in normal dispersion Si$_3$N$_4$ microcavities \cite{Bowers_NP2021hertz,lihachev2022platicon}, the AlN microcomb is associated with lines in the Raman band. Such a state is similar to bright Stokes solitons in amorphous silica microcavities \cite{Vahala_NP2017Stokes}. The microcomb spectrum also agrees well with the simulation based on our nonlinear SIL model augmented by a narrow Raman gain spectrum (see Supplementary Sec. 7 \cite{Supplement}). 
The inset of Fig. \ref{fig4}(a) shows the simulated spectral dynamics in the formation of this state. It suggests the SIL Stokes platicon microcomb forms spontaneously without frequency sweeping, distinct from conventional pumping schemes \cite{Weiner_NP2015mode}.

A platicon microcomb without lines in the Stokes band was also observed by increasing the gap between the chip and the DFB laser (i.e., increasing $\phi_{\rm B}$) in both experiment and simulation (Fig. \ref{fig4}(b)). 
The comb lines in the Stokes band result in temporal oscillations in the platicon as the simulation shows in Fig. \ref{fig4}(c). In the absence of these lines, the oscillations vanish and a typical platicon arises. These microcombs constitute the first demonstrated coherent microcombs in the normal dispersion regime for AlN microcavities (including conventional pumping schemes), to our knowledge. 

 
\textit{Conclusion.-} 
In conclusion, our work establishes the hybrid-integrated AlN nonlinear photonic platform and adds Raman actions to the toolbox of nonlinear SIL systems. SIL-SRL can provide an efficient approach to generate low noise on-chip lasers in the 1700 nm (even the 2000 nm) band. We also elucidate how Raman gain, thermal effects and Kerr effect add functionalities to integrated laser and microcombs via SIL. The SIL-SRL frequency can be swept by several GHz through thermal effects, while $\phi_{\rm B}$ in SIL facilitates platicon state switching via the interplay of Raman and Kerr effects. 
As a crystalline material with an ultrabroad transparency window, the hybrid-integrated AlN platform can offer great opportunities for visible and mid-infrared photonics. 

This work is supported by the National Natural Science Foundation of China (62250071, 62175127, 62375150), by the National Key Research and Development Program
of China (2021YFB2801200, 2023YFB3211200), and by the Tsinghua-Toyota Joint Research Fund.

\bibliography{main.bib}

\end{document}